\documentclass[nofootinbib,showpacs,prd]{revtex4}%
\usepackage{amsmath}
\usepackage{amsfonts}
\usepackage{amssymb}

\def\beqa{\begin{eqnarray}}
\def\eeqa{\end{eqnarray}}
\def\beq{\begin{equation}}
\def\eeq{\end{equation}}

\begin{document}

\title{Tomographic Representation of Minisuperspace Quantum Cosmology  and Noether Symmetries }

\author{S. Capozziello$^{1}$,
V.I. Man'ko$^{1,2}$, G. Marmo$^{1}$, C. Stornaiolo$^{1}$}

\vspace{2mm}

 \affiliation{$^1$Dipartimento di Scienze Fisiche,
Universit\`a di Napoli "Federico II", and INFN, Sez. di Napoli,
Compl. Univ. di Monte S. Angelo, Edificio G, Via Cinthia, I-80126
- Napoli, Italy\\
$^2$ P.N. Lebedev Physical Institute, Leninskii Pr. 53, Moscow
117924, Russia}

\begin{abstract}
The probability representation, in which  cosmological quantum
states  are  described by a standard positive probability
distribution, is constructed for minisuperspace models selected by
Noether symmetries.  In such a case, the tomographic probability
distribution provides the classical evolution for the models and
can be considered an approach to select "observable" universes.
Some specific examples, derived from Extended Theories of Gravity,
are worked out. We discuss also how to connect tomograms,
symmetries and cosmological parameters.
\end{abstract}


\maketitle


\section{Introduction}\label{introduction}

The main issue of Quantum Cosmology is to find out methods,
derived from some quantum gravity approach, to achieve the so
called Wave Function of the Universe (or the corresponding density
matrix), a quantity related to the initial conditions from which
dynamical systems (representing "classical universes") can emerge
\cite{Hartle:1986gn,hartlebru}.

Different representations of the density matrix such as the
 Wigner quasi-distribution were applied  to study cosmological
 models in the minisuperspace \cite{habib,habib1}. Recently, a tomographic
 probability representation of  quantum states was suggested:
 the quantum state can be associated to a standard
 positive probability distribution (tomogram). The tomographic
 probability of quantum states was introduced in Quantum Cosmology
 in \cite{cosimo1}. In this framework, cosmological models described by the
 solutions of the Wheeler-de Witt equation can be mapped onto
 solutions of the Fokker-Planck-like equations for standard
 probability distributions. The tomographic probabilities are
 connected to the Wigner function of the Universe by the Radon
 integral transform \cite{cosimo1} and by the wave function of
 the Universe with the wave related to the fractional Fourier transform
 \cite{cosimo1,cosimo2}
  considered in quantum mechanics in \cite{mendes,margarita}.

 Some important remarks are in order to point out  why the
tomographic approach is useful  to improve  the Wigner picture. As
we said, the Wigner function \cite{wigner32} can be used to
describe quantum states which are "universes". It has been
introduced in order to deal with mixed quantum states with some
remarkable properties. In contrast to the standard wave function
and density matrix representation, where complex functions are
used, the Wigner function is a real one. It depends on two real
variables, $p$ and $q$, which label a point in the phase space.
The marginals constructed by the Wigner function ${\cal W}(p,q)$
are integrals either over $q$ or over $p$. They are positively
defined probability densities in position and in momentum
respectively. These properties make the Wigner function to be very
similar to the  probability distribution density used to describe
the classical states of a particle in the phase space as in
classical statistical mechanics. The Wigner function was
introduced, namely, to find a rigorous description of quantum
states in terms of objects which have a classical probability
distribution. However, such a function cannot play this role since
it can take negative values and this feature does not allow to
interpret the function as a fair probability distribution density.
The necessity to find out such a kind of  probability density,
which rigorously describes the quantum state, is known as the {\it
Pauli problem} \cite{pauli}. Pauli supposed that the probability
distribution in momentum, together with the one in position,
determines completely the quantum state. But it was shown  that
different quantum states can exist  having the same  probability
distribution density, so the Pauli problem was proved to have no
solution, if one has only these two probability distributions
(namely the one in momentum and the one in position).
Nevertheless, this problem got a complete solution in the
tomographic probability approach to quantum state description. In
short, one needs a family of marginal probability densities in the
phase space  obtained from rotations in phase space of the Pauli
distribution pair: such rotations can be parameterized by a given
angle $\theta$. This family, which is nothing else but the Radon
integral transform \cite{Radon1917} of the Wigner function, is
sufficient to completely describe the quantum state, i.e. to find
the density matrix of Wigner function if one knows the tomographic
probability distribution. The tomographic probability contains the
same information on the quantum state as the Wigner function of
density matrix. On the other hand,  like the Wigner function,
being completely equivalent to density matrix in position and
momentum representations, the tomographic probability density has
its own merit:  it is a fair "positive" probability distribution.

It is well known that the Wigner function has been intensively
used in Quantum Cosmology \cite{WignerQC}. Besides, some
applications of tomographic probability, based on its property to
be the fair positive measurable probability distribution, have
been considered for cosmological problems \cite{cosimo1,cosimo2}.
For example, the notion of tomographic entropy, using the Shannon
construction of entropy and information \cite{Shannon1947},
provides extra characteristics of cosmological quantum states
\cite{cosimo1,cosimo2}. The Shannon  entropy cannot be used for
the Wigner function since such a  function is not a probability
distribution.

Furthermore, using the properties of tomographic probability,
related to its own features, it is possible to address several
problems of Quantum Cosmology whose solutions cannot be achieved
by  the Wigner function. In particular, it has been shown that
symmetries can play an important role to select viable
cosmological models \cite{marek,cqg,cimento}. In the framework of
the so-called {\it Noether Symmetry  Approach}
\cite{cimento,marmo,arturo} it is possible, in principle, to find
out cyclic variables related to conserved quantities and then to
reduce cosmological dynamics. Besides, the existence of symmetries
fixes the forms of  couplings and self-interaction potential
giving the relation between them in the interaction Lagrangians.
The existence of Noether symmetries for minisuperspace
cosmological models can be viewed as a sort of selection rule to
recover classical behaviors in cosmic evolution
\cite{Capozziello:1999xr}. In fact, it  was shown in
\cite{Hartle:1986gn} that the form of the wave function with
specific picks provides a classical regime in the evolution of the
Universe. The picks (as oscillations) are characteristic
properties of such regime and are selected in cosmological models
where Noether symmetries are present \cite{Capozziello:1999xr}.

In the framework of minisuperspace approach to Quantum Cosmology,
the aim of this paper is to study cosmological models in order to
seek for conditions to select classical (observable) behaviors by
tomographic probability representation. We find out that,
according to \cite{Capozziello:1999xr}, if tomograms are related
to extra symmetries (Noether symmetries), they allow oscillatory
behaviours of the wave function and then give rise to classical
solutions of dynamics. In the framework of the Hartle criterion,
such solutions are observable universes. The approach is worked
out for general classes of Extended Theories of Gravity (see for
example \cite{garattini,odintsov1} which have acquired  a huge
interest in recent years as possible schemes capable of explaining
cosmological dynamics  at all epochs, starting from inflation,
through matter dominated era, up to the today observed accelerated
behavior \cite{eos,cnot}.

The paper is organized as follows. In Sec.2, the tomographic
approach and its relation to the Wave Function of the Universe,
considering the Wigner function and density is reviewed.  Sec.3 is
devoted to the semiclassical limit of Quantum Cosmology and the
Hartle criterion to select observable universes. In Sec.4, the
stationary phase method is discussed in view of  the WKB
approximation for the Wave Function of the Universe.   Sec.5 is a
short summary of the minisuperspace cosmological models, coming
from Extended Theories of Gravity, which we are going to analyze
in the tomographic representation. Exact solutions (wave function
of the universe) are obtained if Noether symmetries exist. Such
solutions show oscillatory behaviors and then the possibility to
implement the Hartle criterion. In Sec.6, we give the related
tomograms in stationary phase approximation for the previous
selected solutions. Conclusions and a  discussion of possible
relations to the observed universe the classical regime evolution
related to initial are presented in Sec.7.

\section{Tomograms of cosmological quantum states}

We consider here the tomographic map of the wave function of the
universe (or its density matrix) in the framework of the
minisuperspace approach. Let us take into account first the
minisuperspace model in which the pure space of the universe is
described by a vector $|\Psi>|$ in the Hilbert space of quantum
states. The wave function $\Psi(x)=<x|\Psi>$, $x\in \Re$ can be
mapped onto a fair probability distribution ${\cal W}(X,\mu,\nu)$
called tomogram
\begin{equation}\label{univwavfun}
 {\cal W}(X,\mu,\nu)=\frac{1}{2\pi|\nu|}\left|\int\Psi(y)e^{\frac{i\mu}{2\nu}y^{2}-
 \frac{iX}{\nu}y}dy\right|^{2}.
\end{equation}
This probability distribution is positive and normalized for all
the parameters $\mu,\, \nu\, \in \Re$
\begin{equation}\label{4}
 \int{\cal W} (X,\mu,\nu)dX=1,
\end{equation}
if the wave function is normalized, i.e.
\begin{equation}\label{normalized}
\int|\Psi(y)|^{2}=1
\end{equation}
The tomographic map of the wave function of the universe onto the
universe tomogram
\begin{equation}\label{tomomap}
     \Psi(x)\mapsto{\cal W} (X,\mu,\nu)
\end{equation}
is invertible (up to a constant phase factor). One has the inverse
formula determining the density matrix of the universe  pure state
in the positive representation

\begin{equation}\label{pure state}
     \Psi(x)\Psi^{*}(x')=\frac{1}{2\pi}\int {\cal W} (y,\mu,
     x-x')e^{i\left( y-\mu (x+x')/ 2 \right)}dy\,d\mu.
\end{equation}
Thus one has the relation

\begin{equation}\label{relationwft}
     |\Psi(0)|^{2}=\frac{1}{2\pi}\int {\cal W} (y,\mu,
     0)e^{i y}dy\,d\mu
     \end{equation}
In view of Eq.(\ref{pure state}), one has the connection of the
wave function of the universe with its tomogram
\begin{equation}\label{connection}
\Psi(x)=\frac{1}{2\pi|\Psi(0)|}\int {\cal W} (y,\mu,
     x)e^{i\left( y-\mu x/2\right)}dy\,d\mu
\end{equation}
or in view of Eq.(\ref{relationwft})

\begin{equation}\label{rel}
\Psi(x)=\frac{1}{2\pi}\int {\cal W} (y,\mu,
     x)e^{i\left( y-\mu x/2\right)}dy\,d\mu\left[\frac{1}{2\pi}\int {\cal W} (y',\mu',
     0)e^{i y'}dy'\,d\mu'\right]^{-1/2}
\end{equation}
Eq.(\ref{rel}) determines the wave function of the universe in
terms of the universe tomogram (up to a constant phase factor).
This constant phase factor of the wave function is not essential.
In fact, the wave function can be written in form of a modulus and
a phase factor
\begin{equation}\label{decomposition}
    \Psi(x)=|\Psi(x)|e^{iS(x)}.
\end{equation}
In view of Eq.(\ref{rel}) one can obtain both the amplitude and
the phase of the wave function in terms of the tomogram. For the
amplitude, one has
\begin{equation}\label{amplitude}
    |\Psi(x)|^{2}=\frac{1}{2\pi}\int {\cal W} (y,\mu,
     x)e^{i\left( y-\mu x/ 2 \right)}dy\,d\mu.
\end{equation}
Since the amplitude squared is a real (positive) function, it is
expressed in terms of the real part of the right hand side of
Eq.(\ref{amplitude}), i.e.
\begin{equation}\label{amplitude1}
    |\Psi(x)|^{2}=\frac{1}{2\pi}\int {\cal W} (y,\mu,
     x)\cos \left( y-\mu x/ 2 \right)dy\,d\mu.
\end{equation}
We point out that the tomogram ${\cal W} (y,\mu,x)$ is a real non
negative function. On the other hand, the phase of the wave
function  $S(x)$ is also determined by the universe tomogram since
\begin{equation}\label{sofx}
    S(x)=-i\ln \left[\frac{\Psi(x)}{|\Psi(x)|}\right]
\end{equation}
one has
\begin{equation}\label{sofx2}
    S(x)=-i\ln \left[\frac{\int {\cal W} (y,\mu,
     x)e^{i\left( y-\mu x/2\right)}dy\,d\mu}{|\int {\cal W} (y,\mu,
     x)e^{i\left( y-\mu x/2\right)}dy\,d\mu|}\right].
\end{equation}
In quasi classical approximation (WKB) for the wave function of
the universe, which corresponds to a stationary  state with energy
$E$, the amplitude and the phase are connected. In this case the
phase $S(x)$ satisfies the Hamilton-Jacobi equation. One can
consider the tomogram given by Eq.(\ref{univwavfun}) as follows
\begin{equation}\label{univwavfunwkb}
 {\cal W}(X,\mu,\nu)=\frac{1}{2\pi|\nu|}\left|\int|\Psi(y)|e^{i S(y)+\frac{i\mu}{2\nu}y^{2}-
 \frac{iX}{\nu}y}dy\right|^{2}.
\end{equation}
The approach outlined above is in complete agreement with the
semiclassical limit of Quantum Cosmology and it provides a
physical interpretation of the wave function of the universe, as
we are going to discuss below.

\section{The semiclassical limit of Quantum Cosmology and the
Hartle Criterion}

Semiclassical limit occupies an important role in Quantum Gravity
due to the lack of a self-consistent and definitive theory. Only
thanks to such a limit, we can obtain the solution of the
Wheeler-De Witt (WDW) equation, the wave function of the universe
$\Psi$, and search for interpretative  criteria which allow
physically motivated previsions. Furthermore, cosmological
observations are based on a ``classical universe'' since we are
not able, at least considering today's facilities, to obtain data
toward and before the Planck epoch. Taking into account also the
fact that the main goal of Quantum Cosmology is to give
self-consistent laws for initial conditions, the correct
identification of a semiclassical limit is crucial. In standard
quantum mechanics, the wave function can be developed as a power
law series in $\hbar$ in the WKB approximation. Formally, the
semiclassical limit corresponds to the  limit $\hbar\to 0$.
Analogously, we can write down the WDW equation as

\begin{equation}\label{wdwequation}
     \left(\frac{1}{m_{P}^{2}}\nabla^{2}-m_{P}^{2}
     U\right)\Psi[h_{ij}(x), \phi(x)]=0
\end{equation}
where $U(h_{ij},\phi)$ is the superpotential, $h_{ij}$ are the
components of the spatial three-metric, the geometrodynamical
variables, and $\phi$ is a generic scalar field describing the
matter content. The semiclassical limit can be achieved by using
$m_{P}^{-2}$, the squared Planck mass, as an expansion parameter.

The wave function can be represented as
\begin{equation}\label{wavefunction}
  \Psi[h_{ij}(x), \phi(x)]\sim e^{im_{P}^{2} S}
\end{equation}
where $S$ is the an action. A state with classical correlations
has to be a superposition of states  of the form
(\ref{wavefunction}). The WKB approximation is achieved by the
expansion
\begin{equation}\label{expansion}
    S= S_0 + m_{P}^{-2}S_{1}+ O(m_{P}^{-4})
\end{equation}
and inserting it into the WDW equation. Then equating terms of the
same order in $m_{P}$, we obtain the Hamilton-Jacobi equation at
any order  in $S$. At the lowest order in $S_{0}$, we get
\begin{equation}\label{HJordinezero}
    \nabla S_{0}\cdot\nabla S_{0} + U =0\,.
\end{equation}
In the same way, we obtain equations at higher orders which can be
solved taking into account results from the previous orders. It
can be shown that we need only $\nabla S_{0}$ to recover the
semi-classical limit of Quantum Cosmology \cite{halliwell}, that
is the wave function of the Universe is given by
\begin{equation}\label{wavefunctionsemiclass}
  \Psi \sim e^{im_{P}^{2} S_{0}}\,.
\end{equation}
If $S_{0}$ is a real number, we have  oscillating WKB modes and
$\Psi$ is peaked on a phase-space region defined by the equations
\begin{equation}\label{momentah}
     \Pi_{ij}=m_{P}^{2}\frac{\delta S_{0}}{\delta h^{ij}},
\end{equation}
\begin{equation}\label{momentafi}
    \Pi_{\phi}=m_{P}^{2}\frac{\delta S_{0}}{\delta \phi}
\end{equation}
where  $\Pi_{ij}$ and $\Pi_{\phi}$ are respectively  the classical
 momenta conjugate to $h^{ij}$ and $\phi$.

 Using the Hamilton-Jacobi Eq.(\ref{HJordinezero}),
 Eqs.(\ref{momentah}) and (\ref{momentafi}) are first
 integrals  of classical field equations defining a set of
 solutions. This fact, in principle, can solve the problem of
 initial conditions. In fact, for a given action $S_{0}$,
 solutions (\ref{momentah}) and (\ref{momentafi}) involve $n$
 arbitrary parameters, while the general solution of the field
 equations involves $\{2n-1\}$ parameters. Then the wave function is
 peaked (or oscillates, being $S_{0}$ real) on a subset of the
 general solution. In other words, the boundary conditions on the
 wave function imply initial conditions for the classical
 solution. Besides, if the wave function $\Psi$ is sufficiently
 peaked, oscillating about some region of the configuration space
 $Q=\{h^{ij},\phi\}$, it is possible  to find correlations among the observables
 that characterize this region.
 If it is not peaked, correlations among the observables are not
 present.

 At this point a discussion on what we mean as ``classical
 universe'' and how the concept is related to an oscillating
 $\Psi$ is due.

Let us take as an example a cosmological model characterized by a
set of observable parameters as $H_{0}$, the Hubble constant,
$q_{0}$, the deceleration parameter, $\Omega_{M}$, the density
parameter for matter, $\Omega_{\Lambda}$ the density parameter for
the cosmological constant $\Lambda$ (or dark energy) and
$\Omega_{k}$, the amount of spatial curvature \footnote{Actually,
one should consider also the age of the universe $t_0$, but,
usually, it is  referred as $t_0\sim H_0^{-1}$.}. All of them are
related by some cosmological model, solution of the
Einstein-Friedmann equations and can be measured with a certain
accuracy, by observations (see e.g. \cite{IJGMMP} and references
therein for a discussion). A ``good'' wave function of the
universe is that one which is ``peaked'' about such a solution of
the cosmological equations (which can be alternatively derived
from the momenta (\ref{momentah}) and (\ref{momentafi}) and then
``correlates'' the set of observables
$\{H_{0},q_{0},\Omega_{M},\Omega_{\Lambda},\Omega_{k} \}$. It is
worth noticing that $\Psi$  does not predict any specific value of
the set of observables, but guarantees the correlations among
them. This is the so-called {\it Hartle criterion}
\cite{Hartle:1986gn}. On the other hand, such a correlation allows
to define a parameter able to label the points along the classical
trajectories, solutions of (\ref{momentah}) and (\ref{momentafi}).
Defining a tangent vector in the configuration space of classical
trajectories, where $\Psi$ is peaked, it is
\begin{equation}\label{tangentvector}
     \frac{d}{d\tau}= 2 \nabla S_{0}\cdot \nabla
\end{equation}
where $\tau$ is the ``proper time'' along the classical
trajectories. From these considerations, we can state that
\begin{enumerate}
    \item time emerges as a parameter which labels points along
    trajectories where the wave function is peaked;
    \item time parametrization invariance (a symmetry of classical
    theory) comes out as the freedom to choose such a parameter.
    For example, assuming the redshift $z$, in principle,
    every classical cosmological solution can be expressed in terms
    of $z$;
    \item classical time, and in general, classical spacetime are
    proper notions which emerge only in a superspace regions where
    the wave function is oscillating;
    \item the existence of oscillating wave function, and then of
    classical spacetimes, depends on the boundary conditions of
    the theory, in particular on the shape of superpotential
    $U(h_{ij},\phi)$.
\end{enumerate}

At this point it is clear  what is the semiclassical limit of the
theory. It is  the superspace region where the wave function
oscillates with big values of the phase which indicates strong
correlation among the dynamical variables of the form
(\ref{momentah}) and (\ref{momentafi}). Such variables describe
the classical trajectories in the superspace and make the concept
of classical spacetime emerge. The tangent vector to this
classical paths  defines the proper time. Formally, in analogy
with standard quantum mechanics, the semiclassical limit is
obtained  for $m_{P}\to \infty$, which means low energies  with
respect to the Planck scale. Furthermore, the limiting conditions
for the wave function select a particular set of  classical
universes and the ``measure'' defined by the $\Psi$ itself
indicates a ``typical'' universe. In principle, this is the way in
which Quantum Cosmology approaches the problem of initial
conditions.

The semiclassical region of superspace, defined by the oscillating
structure of this wave function, is called the ``Lorentzian''
region. This can be identified as the `` ensemble'' of all
3-geometries embedded in a classical spacetime. The region outside
the Lorentzian one is called ``Euclidean''. Here the action is
imaginary, i.e. $S_{0}=i I$ and the wave function is exponential
\begin{equation}\label{8}
    \Psi\sim e^{-m_{P}^{-2} I}.
\end{equation}

The wave function of this form is not ``classical'' since it
corresponds to an Euclidean spacetime.

If $\Psi$ is a WKB solution , $I$ is the action of Euclidean
solutions of field equations called ``instantons''. Unlike the
Lorentzian case, the wave function (\ref{8}) is not peaked on a
set of instantons. It is not ``classical'' since it cannot predict
classical correlations among Lorentzian momenta and their
conjugate variables. With these considerations in mind, it is
clear that any method by which conserved quantities, corresponding
to conserved momenta, can be achieved is useful to define a
semiclassical limit in Quantum Cosmology. I other words, conserved
momenta like $\Pi_{ij}=\Sigma_0$ or $\Pi_{\phi}=\Sigma_1$ allow,
in principle, to select classical trajectories corresponding to
observable universes.

In general, the Hamiltonian constraint gives the WDW equation, so
that if $\vert \Psi>$ is a {\it state} of the system (i.e. the
wave function of the universe), dynamics is given by
 \beq\label{39}
 {\cal H}\vert\Psi>=0\,{.}
 \eeq
In \cite{Capozziello:1999xr}, it is shown that if  Noether
symmetries exist, a reduction procedure of dynamics can be
implemented and then,  we get
 \beqa
 -i\partial_1\vert\Psi>&=&\Sigma_1\vert\Psi>\,{,} \nonumber \\
 -i\partial_2\vert\Psi>&=&\Sigma_2\vert\Psi>\,{,} \label{41} \\
 \ldots & & \ldots \,{,} \nonumber
 \eeqa
which are  translations along the $q^j$ axes of configuration
space (minisuperspace) singled out by the corresponding symmetry.
Eqs. (\ref{41}) can be immediately integrated and, being
$\Sigma_j$ real constants, we obtain oscillatory behaviors for
$\vert \Psi>$ in the directions of symmetries, i.e.
 \beq\label{42}
 \vert\Psi>=\sum_{j=1}^m\, e^{i\Sigma_jQ^j}\vert \chi(Q^l)>\,{,}
 \quad m < l\leq n\,{,}
 \eeq
 where $m$ is the number of symmetries, $l$ are the directions
where symmetries do not exist, $n$ is the total dimension of
minisuperspace. Viceversa, dynamics given by (\ref{39}) can be
reduced by (\ref{41}) if and only if it is possible to define
constant conjugate momenta, that is oscillatory behaviors of a
subset of solutions $\vert\Psi>$ exist only if Noether symmetry
exists for dynamics.

The $m$ symmetries give first integrals of motion and then the
possibility to select classical trajectories. In one and
two-dimensional minisuperspaces, the existence of a Noether
symmetry allows the complete solution of the problem and to get
the full semi-classical limit of Quantum Cosmology. As we shall
see below, the tomographic probability representation gives
results strictly related to this method since quantum tomograms
become "classical" as soon as Noether symmetries exist. This
occurrence allows, in principle, to trace back the cosmic
evolution from quantum to classically observable states for
several Extended Theories of Gravity.

It is worth noticing that the present approach has to be compared
with previous results where minisuperspace method has been applied
to similar models. For example, in \cite{kiefer}, it is studied
the quantum-to-classical transition in Jordan-Brans-Dicke quantum
gravity showing the relevance of the method for inflation. In
\cite{orfeu1,orfeu2} the scale factor duality and the cosmological
constant problem are faced from the same minisuperspace point of
view. As we are going to demonstrate, the tomographic approach is
fully coherent with these results.

\section{The Stationary Phase Method}\label{spm}

Let us now evaluate the integral (\ref{univwavfunwkb}) using the
stationary phase approximation in view to select classical state
by tomographic representation. The method is connected with the
evaluation of an integral of the form

\begin{equation}\label{ph0}
  I=   \oint A(z)e^{i\Phi(z)}dz
\end{equation}
where $z$ is a complex variable and a series decomposition of the
function $\Phi(z)$ is used near its extremum
\begin{equation}\label{ph1}
     \Phi(z)\approx \Phi(z_{0})+\frac{1}{2}\frac{\partial^{2}\Phi}{\partial
     z^{2}}\Big|_{z=z_{0}}(z-z_{0})^{2}.
\end{equation}
The point(s) $z_{0}$ is(are) the solutions of the equation
\begin{equation}\label{ph2}
     \frac{\partial\Phi}{\partial z}(z_{0})=0\,.
\end{equation}
The function $A(z)$ is considered  slowly varying near the point
$z_{0}$. Then the integral (\ref{ph0}) reads
\begin{equation}\label{ph3}
    I\approx
    A(z_{0})\frac{\sqrt{\pi}}{\sqrt{-\frac{i}{2}\Phi^{''}(z_{0})}}e^{i\Phi(z_{0})}\,.
\end{equation}
In the case of  $N$  points $z_{0}\rightarrow z_{0}^{(k)}$ (with
$k=1,\dots N$),  the integral assumes the form of a sum over these
points.

Let us consider, in an explicit exact expression
(\ref{univwavfunwkb}), the modulus $|\Psi(y)|$ to be a slowly
varying amplitude function and the function
\begin{equation}\label{ph4}
    \Phi(q)=S(q)+\frac{\mu q^{2}}{2\nu}- \frac{X q}{\nu}.
\end{equation}
The point $y_{0}$ is determined by the expression
\begin{equation}\label{ph5}
    \frac{\partial S}{\partial q} + \frac{\mu}{\nu} q -
    \frac{X}{\nu}=0\,.
\end{equation}
One can see that using the notation
\begin{equation}\label{ph6}
    \frac{\partial S}{\partial q}=p
\end{equation}
the relation (\ref{ph5}) is equivalent to
\begin{equation}\label{ph7}
    \mu q + \nu p = X
\end{equation}
One can remind that in classical mechanics the tomogram is defined
as \cite{lecture}
\begin{equation}\label{ph8} {\cal W}
(X,\mu,\nu)=\int f(q,p)\, \delta(X- \mu q - \nu p)dq\,dp.
\end{equation}
In quantum mechanics, the tomogram is determined via the Wigner
function $ W(q,p)$ as \cite{mancini}
\begin{equation}\label{ph9}
{\cal W} (X,\mu,\nu)=\int W(q,p)\, \delta(X- \mu q - \nu
p)\frac{dq\,dp}{2\pi}.
\end{equation}
Both relations mean the validity of (\ref{ph7}). The Wigner
function is determined by the wave function \cite{wigner32}
\begin{equation}\label{ph10}
     W(q,p)=\int\Psi\left(q+\frac{u}{2}\right)\Psi^{*}\left(q-\frac{u}{2}\right)e^{-iup}du.
\end{equation}
In WKB approximation, the expression for the Wigner function is
given just by the stationary phase method of evaluating the
integral (\ref{ph10}) that is
\begin{equation}\label{ph11}
     W(q,p)=\int\left|\Psi\left(q+\frac{u}{2}\right)\right|\,\left|\Psi^{*}\left(q-\frac{u}{2}\right)\right|
     e^{iS\left(q+\frac{u}{2}\right)}e^{-iS\left(q-\frac{u}{2}\right)}e^{-iup}du.
\end{equation}
This expression was studied and applied in the cosmological
context \cite{habib,habib1}.

 Since the Wigner function
determines the tomograms by (\ref{ph9}), one can get the quasi
classical approximation for the tomogram inserting into
(\ref{ph9}) the Berry's expression for the Wigner function. But as
it is clear, one can get the approximation directly using in
(\ref{ph3}) expression the function $\Phi(q)$ given by
(\ref{ph4}). Then the point $q_{0}$ is defined by (\ref{ph5}) and
$\Phi^{''}(q_{0})$ reads
\begin{equation}\label{ph12}
    \Phi^{''}(q_{0})=\frac{\partial^{2} S(q)}{\partial
    q^{2}}\Big|_{q=q_{0}} + \frac{\mu}{\nu}.
\end{equation}
Thus one has
\begin{equation}\label{ph13}
   {\cal W} (X,\mu,\nu)\approx \frac{1}{|\nu|}\left| \Psi(q_{0})\right|^{2}\left| \frac{\partial^{2} S}{\partial
    q^{2}}(q_{0}) + \frac{\mu}{\nu}\right|^{-1}=
    \frac{\left| \Psi(q_{0})\right|^{2}}{\left| \frac{\partial^{2} S}{\partial
    q^{2}}(q_{0})\nu +\mu\right|}
\end{equation}
More generally, if  Eq.(\ref{ph5}) has more solutions
$q_{0}^{(1)}\dots q_{0}^{(N)}$, the tomogram takes the form
\begin{equation}\label{ph14}
   {\cal W} (X,\mu,\nu)\approx \left|\sum_{i=1}^{N}\frac{ \Psi(q_{0}^{(i)})}{\sqrt{\left| \frac{\partial^{2} S}{\partial
    q^{2}}\Big|_{ q_{0}^{(i)}}\nu +\mu\right|}}\right|^{2}.
\end{equation}
The relation with the semiclassical limit of quantum cosmology is
straightforward. In fact, the existence of  such stationary points
means that the wave function of the universe is peaked on
conserved momenta and then classical trajectories in the
minisuperspace can be found out.

\section{Minisuperspace models from Extended Theories of Gravity}
The previous discussion  can be applied to several classes of
minisuperspaces, in particular one can take in to account Extended
Theories of Gravity as the scalar-tensor  theories or higher-order
theories of gravity. Such theories are interesting since they are
directly related to the issue of recovering  suitable effective
actions in  quantum gravity \cite{odintsov}. Starting from
pioneering works of Sakharov \cite{sakharov}, the effects of
vacuum polarization on the gravitational constant, i.e. the fact
that gravitational constant can be induced by vacuum polarization,
have been extensively investigated. All these attempts led to take
into account gravitational actions extended beyond the simple
Hilbert-Einstein action of General Relativity (GR) which is linear
in the Ricci scalar $R$. The motivation is to investigate
alternative theories in order to  cure the shortcomings of GR,
essentially due to the emergence of singularities in
high-curvature regimes.

The Brans--Dicke approach is one of this attempt which, asking for
dynamically inducing the gravitational coupling by a scalar field,
is more coherent with the Mach principle requests \cite{brans}.

Besides, it has been realized that corrective terms are
inescapable if we want to obtain the effective action of quantum
gravity at scales close to the Planck length (see e.g.
\cite{vilkovisky}). In other words, it seems that, in order to
construct a renormalizable theory of gravity, we need higher-order
terms of curvature invariants such as $R^2, R^{\mu\nu}R_{\mu\nu},
R^{\mu\nu\alpha\beta}R_{\mu\nu\alpha\beta}, R\Box R$, $R\Box^kR$
or nonminimally coupled terms between scalar fields
 and geometry as $\varphi^2R$ (see \cite{francaviglia1,nojiri,francaviglia} for a review).

Stelle \cite{stelle} constructed a renormalizable theory of
gravity by introducing quadratic terms in curvature invariants.
Barth and Christensen gave a detailed analysis of the one-loop
divergences of fourth-order gravity theories providing the first
general scheme of quantization of higher-order theories
\cite{christensen,barth}. Several results followed  and today it
is well known that a renormalizable theory of gravity is obtained,
at least at one-loop level, if  quadratic terms in the Riemann
curvature tensor and its contractions are introduced
\cite{odintsov}. Any action, where a finite number of terms
involving power laws of curvature tensor or its derivatives
appears, is a low-energy approximation to some fundamental theory
of gravity which, up to now, is not available. For example, String
Theory or Supergravity present low-energy effective actions where
higher-order or nonminimally coupled terms appear \cite{fradkin}.

However, if Lagrangians with higher-order terms or arbitrary
derivatives in curvature invariants are considered, they are
expected to be non-local and give rise to some characteristic
length $l_0$ of the order of Planck length. The expansion in terms
of $R$ and $\Box R$, for example, at scales larger than $l_0$
produces infinite series which should break near $l_0$
\cite{birrel}. With these facts in mind, taking into account such
Lagrangians, means to make further steps toward a complete
renormalizable theory of gravity. Cosmological models coming from
such effective theories have been extensively studied by the
 {\it Noether Symmetry Approach}. In particular, nonminimally coupled theories of the
form
 \begin{equation}\label{eq4}
 {\cal L}=\sqrt{-g}\,\left[F(\varphi)R+\frac{1}{2}\nabla_{\mu}\varphi
 \nabla^{\mu}\varphi-V(\varphi)\right]\,{,}
 \end{equation}
 where $ F(\varphi)$ and $V(\varphi)$ are
respectively the coupling and the potential of the scalar field,
and fourth-order theories like
 \begin{equation}\label{equafor}
 {\cal L}=\sqrt{-g}\,f(R)\,,
 \end{equation}
where $f(R)$ is a generic function of the scalar
curvature\footnote{The field equations of $f(R)$ theories are  of
fourth-order in metric derivatives. They reduce to the standard
second order Einstein equations for $f(R)=R$.}. In
\cite{marek,cqg,arturo,gaetanohigh}, it was shown that asking for
the existence of a Noether symmetry
 \begin{equation}\label{eq5}
  L_{X}{\cal L}=0\to X{\cal L}=0\,{,}
 \end{equation}
where $ L_{X}$ is the Lie derivative with respect to the Noether
vector $X$, it is possible to select physically interesting forms
of the interaction potential $V(\varphi)$, the gravitational
coupling $F(\varphi)$ and the function $f(R)$.

As we said above, the existence of Noether symmetries allows to
select constants of motion so that the dynamics results
simplified. Often such a dynamics is exactly solvable by a
straightforward change of variables where a cyclic one is present.
This occurrence reveals extremely useful in Quantum Cosmology
since allows to formulate minisuperspace models exactly solvable.
Considering the above cases, i.e. scalar-tensor and fourth-order
gravity,  suitable 2-dimensional configuration spaces
(minisuperspace) can be achieved adopting a
Friedmann-Robertson-Walker (FRW) metric which gives the pointlike
Lagrangians.

In the case of scalar-tensor theories, we have
\begin{equation}\label{47}
 {\cal L}=6a\dot{a}^2F+6a^2\dot{a}\dot{F}-
 6kaF+a^3\left[\frac{\dot{\varphi}}{2}-V\right]\,{,}
 \end{equation}
 in terms of the scale factor $a$.

The configuration space of such a Lagrangian is ${\cal
Q}\equiv\{a, \varphi\}$, i.e. a two--dimensional minisuperspace. A
Noether symmetry exists if (\ref{eq5}) holds. The related Noether
vector has to be
 \begin{equation}\label{48}
 X=\alpha\, \frac{\partial}{\partial a}+
 \beta\,\frac{\partial}{\partial \varphi}+
 \dot{\alpha}\,\frac{\partial}{\partial\dot{a}}+
 \dot{\beta}\,\frac{\partial}{\partial\dot{\varphi}}\,{,}
 \end{equation}
where $\alpha, \beta$ depend on $a, \varphi$. The condition
(\ref{eq5}) gives rise to a set of partial differential equations
whose solutions are $\alpha, \beta$, $F(\varphi)$ and $V(\varphi)$
(see \cite{Capozziello:1999xr} for details). For example,  a
solution is
 \begin{equation}\label{53}
 \alpha=-\frac{2}{3}p(s)\beta_0a^{s+1}\varphi^{m(s)-1}\,{,}\quad
 \beta=\beta_0a^s\varphi^{m(s)}\,{,}
 \end{equation}
 \begin{equation}\label{54}
 F(\varphi)=D(s)\varphi^2\,{,}\quad
 V(\varphi)=\lambda\varphi^{2p(s)}\,{,}
 \end{equation}
 where
 \begin{equation}\label{55}
 D(s)=\frac{(2s+3)^2}{48(s+1)(s+2)}\,{,}\quad
 p(s)=\frac{3(s+1)}{2s+3}\,{,}\quad
 m(s)=\frac{2s^2+6s+3}{2s+3}\,{,}
 \end{equation}
 and $s, \lambda$ are free parameters. A suitable change of variables
 gives
 \begin{equation}\label{56}
 w=\sigma_0a^3\varphi^{2p(s)}\,{,}\quad z=\frac{3}{\beta_0\chi(s)}
 a^{-s}\varphi^{1-m(s)}\,{,}
 \end{equation}
 where $\sigma_0$ is an integration constant and
 \begin{equation}\label{57}
 \chi(s)=-\frac{6s}{2s+3}\,{.}
 \end{equation}
 Lagrangian (\ref{47}) becomes, for $k=0$,
 \begin{equation}\label{58}
 {\cal L}=\gamma(s)w^{s/3}\dot{z}\dot{w}-\lambda w\,{,}
 \end{equation}
 where $z$ is cyclic and
 \begin{equation}\label{59}
 \gamma(s)=\frac{2s+3}{12\sigma_0^2(s+2)(s+1)}\,{.}
 \end{equation}
 The conjugate momenta are
 \begin{equation}\label{60}
 \pi_z=\frac{\partial {\cal L}}{\partial \dot{z}}=\gamma(s)w^{s/3}
 \dot{w}=\Sigma_0\,{,}\qquad
 \pi_w=\frac{\partial {\cal L}}{\partial \dot{w}}=\gamma(s)w^{s/3}
 \dot{z}\,{,}
 \end{equation}
and then the WDW equation is
 \begin{equation}\label{64}
 [(i\partial_z)(i\partial_w)+\tilde{\lambda}w^{1+s/3}]\vert\Psi>=0\,{,}
 \end{equation}
 where $\tilde{\lambda}=\gamma(s)\lambda$.

The quantum version of the first of momenta (\ref{60}) is
 \begin{equation}\label{65}
 -i\partial_z\vert\Psi>=\Sigma_0\vert\Psi>\,{,}
 \end{equation}
 so that dynamics results reduced. A straightforward integration
 of Eqs. (\ref{64}) and (\ref{65}) gives
  \begin{equation}\label{66}
  \vert\Psi>=\vert\Omega(w)>\vert\chi(z)>\propto
  e^{i\Sigma_0z}\,e^{-i\tilde{\lambda}w^{2+s/3}}\,{,}
  \end{equation}
which is an oscillating wave function.

Analogously, in fourth-order gravity case, the pointlike FRW
Lagrangian is
 \begin{equation}\label{77}
 {\cal L}=6a\dot{a}^2p+6a^2\dot{a}\dot{p}-6kap-a^3W(p)\,{,}
 \end{equation}
 which is of the same form of (\ref{47}) a part the kinetic term.,
 being $p=f'(R)$.
This is an Helmhotz-like Lagrangian \cite{magnano} and the
configuration space is now ${\cal Q}\equiv\{a, p\}$; $p$ has the
same role of the above $\varphi$. Condition (\ref{eq5}) is now
realized by the vector field
 \begin{equation}\label{80}
 X=\alpha (a, p)\frac{\partial}{\partial a}+\beta(a, p)
 \frac{\partial}{\partial p}+\dot{\alpha}\frac{\partial}{\partial\dot{a}}
 +\dot{\beta}\frac{\partial}{\partial\dot{p}}
 \end{equation}
The solution of this system, i.e. the existence of a Noether
symmetry, gives $\alpha$, $\beta$ and $W(p)$. It is satisfied for
\begin{equation}\label{85}
  \alpha=\alpha (a)\,{,} \qquad \beta (a, p)=\beta_0 a^sp\,{,}
 \end{equation}
where $s$ is a parameter and $\beta_0$ is an integration constant.
In particular,
 \begin{equation}\label{86}
  s=0\longrightarrow \alpha (a)=-\frac{\beta_0}{3}\, a\,{,} \quad
 \beta (p)=\beta_0\, p\,{,} \quad W(p)=W_0\, p\,{,} \quad
 k=0\,{,}
 \end{equation}
 \begin{equation}\label{87}
 s=-2\longrightarrow \alpha (a)=-\frac{\beta_0}{a}\,{,}\quad
 \beta (a, p) = \beta_0\, \frac{p}{a^2}\,{,} \quad
 W(p)=W_1p^3\,{,} \quad \forall \,\,\, k\,{,}
 \end{equation}
where $W_0$ and $W_1$ are constants. As above, the new set of
variables ${\cal Q}\equiv\{z, w\}$ adapted to the foliation
induced by $X$ can be achieved. The procedure to achieve the
solution for the WDW equation is exactly the same. The results are
summarized in Tab.1.
\begin{table}
 \begin{tabular}{|c|c|c|c|}
   \hline
 &   &   &   \\
  Gravitational theory &$p$ & $z$ & $w$  \\
    &   &   &   \\  \hline
      &   &   &   \\
   Scalar-Tensor  & $\frac{(2s+3)^2}{48(s+1)(s+2)}\varphi^2$ &  $\frac{3}{\beta_{0}\chi(s)}a^{-5}\varphi^{1-m(s)}$ &
   $\sigma_{0}a^{3}\varphi^{2l(s)}$
   \\
    &   &   &  \\ \hline
     &   &   &   \\
   Fourth order gravity $s=-2$ & $ f'(R) $ & $\ln p $& $ap$ \\
   &   &   &  \\ \hline
    &   &   &   \\
 Fourth order gravity $s=0$ & $f'(R)$ &$ a^{2}$ &$ a^{3} p$ .
 \\  &   &   &  \\  \hline
 %
 \end{tabular}
 \caption{Summary of the changes of variables induced by the
 Noether symmetry for scalar-tensor and fourth-order
 minisuperspace models. The parameters as functions of $s$ are
 $\chi(s)=-6s/(2s+3)$,
$m(s)=(2s^{2}+6s+3)/(2s+3)$ and $l(s)=3(s+1)/(2s+3)$.}
\end{table}
The solutions for the respective WDW equations are given in Tab.2.
\begin{table}

 \begin{tabular}{|c|c|}
   \hline
 &      \\
  Gravitational theory & $\Psi(z,w)$  \\
    &      \\  \hline
      &      \\
   Scalar-Tensor  & $e^{i\Sigma_{0}z}e^{-i\tilde{\lambda}w^{2+s/3}} $
   \\
    &     \\ \hline
     &      \\
   Fourth order gravity $s=-2$ & $e^{i[\Sigma_{1}z + 9 k  w^{2}  + (3w_{1}/4)w^{4}]}  $ \\
   &      \\ \hline
    &       \\
 Fourth order gravity $s=0$ & $ e^{i\Sigma_{0}[z-(1/4)\ln w ]}w^{1/2}Z_{\tilde{\nu}}(\lambda,w)$
 \\  &      \\  \hline
   &       \\
 Fourth order gravity $s=0$ and $w>>0$ & $ e^{i[\Sigma_{0}z-(\Sigma_{0}/4)\ln w \pm \lambda w]}$
 \\  &      \\  \hline
 \end{tabular}
\caption{Summary of the solutions of the WDW equation for the
minisuperspace models. The $Z_{\tilde{\nu}}(\lambda,w)$ are Bessel
function.}
\end{table}
By a rapid inspection of the wave functions in Tab.2, it is clear
that the oscillatory behavior, i.e. the presence of peaks, is
strictly related to the Noether constant $\Sigma_0$. In other
words, the Hartle criterion can be immediately implemented when
conserved quantities are found out.

\section{Minisuperspace tomograms in stationary phase approximation}

The above results can be immediately translated in the tomographic
representation.  If Noether symmetries are present, the solutions
of the WDW equation,  take the following form
\begin{equation}\label{psinoether}
     \Psi(Q_{1},\dots,Q_{n})=\sum_{j=1}^{m}e^{i\Sigma_{j}Q^{j}}\chi
     (Q^{l})\ \ \ \ \ \ \ \ m<l\leq n
\end{equation}
then we find, according to the results of Sec.\ref{spm}, that the
corresponding tomograms are

   $$ \mathcal{W}(X_{1}\dots X_{n},\mu_{1}, \dots ,\mu_{n}, \nu_{1},
    \dots \nu_{n})=\sum_{j=1}^{m} \frac{1}{\mu_{j}}\left|\int\chi (Q^{l})\right.$$
   \begin{equation}\label{tomonoether}
   \times\left.\left(\prod_{l=m+1}^{n} e^{\frac{i\mu}{2\nu}(Q^{l})^{2}-
 \frac{iX}{\nu}(Q^{l})}d(Q^{l})\right)\right|^{2}\,.
\end{equation}
It is  then straightforward to study the form of the tomograms
satisfying  the Hartle criterion.

Let us consider first the tomograms resulting from the
scalar-tensor theories. From  Tab.2, the wave function takes the
form
\begin{equation}\label{st1}
\Psi(z,w) \propto e^{i\Sigma_{0}z}e^{-i\tilde{\lambda}w^{2+s/3}}.
\end{equation}
If we consider the  case $s=0$,  the corresponding two variable
tomogram is
 \begin{equation}\label{st2}
     \mathcal{W} (X_{1},X_{2}, \mu_{1},\mu_{2},\nu_{1},\nu_{2})
     \propto\frac{1}{|\mu_{1}|}\frac{1}{|\mu_{2}-2\tilde{\lambda}\nu_{2}|}.
 \end{equation}
For the  fourth order gravity models, we have considered the two
cases  $s=-2$ and $s=0$. In the first case, the wave function is
\begin{equation}\label{st3}
\Psi(z,w) \propto  e^{i[\Sigma_{1}z + 9 k  w^{2}  +
(3w_{1}/4)w^{4}]}
\end{equation}
and the corresponding tomogram is
\beq\label{st4}
     \mathcal{W} (X_{1},X_{2}, \mu_{1},\mu_{2},\nu_{1},\nu_{2})\propto\frac{1}{|\mu_{1}|}
      \left|\sum_{j=1}^{3}\frac{ \exp(i(9k\alpha_{j}^{2}(X_{2},\mu_{2},\nu_{2})+
      \frac{3w_{1}}{4} \alpha_{j}^{4}(X_{2},\mu_{2},\nu_{2}) )}{\sqrt{9w_{1}\nu_{2}+18k\nu_{2}+\mu_{2}}}\right|^{2}
 \eeq
where $\alpha_{1}$, $\alpha_{2}$, $\alpha_{3}$ are the solutions of
equation
\begin{equation}\label{st5}
      3w_{1}w^{3}+\left(\frac{\mu_{2}}{\nu_{2}}+18k\right)w-
       \frac{X_{2}}{\nu_{2}} =0
\end{equation}

 $$ \alpha_{1}=\frac{-3\cdot2^{ 1/3}\,
     \left( \mu  +
       18\,k\,\nu  \right) }
     {{\left( -81\,X\,
         {\nu }^2 +
        {\sqrt{6561\,X^2\,
           {\nu }^4 +
           2916\,{\nu }^3\,
           {\left( \mu  +
           18\,k\,\nu  \right)
           }^3}} \right) }^
     {1/3}} $$
     $$+
  \frac{{\left( -81\,X\,
         {\nu }^2 +
        {\sqrt{6561\,X^2\,
           {\nu }^4 +
           2916\,{\nu }^3\,
           {\left( \mu  +
           18\,k\,\nu  \right)
           }^3}} \right) }^
     {\frac{1}{3}}}{3\cdot
     2^{1/3}\,\nu },$$

$$\alpha_{2}=\frac{3\,\left( 1 +
       i \,{\sqrt{3}} \right) \,
     \left( \mu  + 18\,k\,\nu
       \right) }{2^{2/3}\,
     {\left( -81\,X\,{\nu }^2 +
         {\sqrt{6561\,X^2\,
            {\nu }^4 +
             2916\,{\nu }^3\,
             {\left( \mu  +
             18\,k\,\nu  \right) }^3
             }} \right) }^{1/3}}
$$
$$
              -
  \frac{\left( 1 -
       i \,{\sqrt{3}} \right) \,
     {\left( -81\,X\,{\nu }^2 +
         {\sqrt{6561\,X^2\,
            {\nu }^4 +
             2916\,{\nu }^3\,
             {\left( \mu  +
             18\,k\,\nu  \right) }^3
             }} \right) }^{1/3}}{6\cdot
     2^{1/3}\,\nu },
     $$
       $$\alpha_{3}=\frac{3\,\left( 1 -
       i\,{\sqrt{3}} \right) \,
     \left( \mu  + 18\,k\,\nu
       \right) }{2^{2/3}\,
     {\left( -81\,X\,{\nu }^2 +
         {\sqrt{6561\,X^2\,
            {\nu }^4 +
             2916\,{\nu }^3\,
             {\left( \mu  +
             18\,k\,\nu  \right) }^3
             }} \right) }^
      {1/3}}
      $$
       $$-
  \frac{\left( 1 +
       i\,{\sqrt{3}} \right) \,
     {\left( -81\,X\,{\nu }^2 +
         {\sqrt{6561\,X^2\,
            {\nu }^4 +
             2916\,{\nu }^3\,
             {\left( \mu  +
             18\,k\,\nu  \right) }^3
             }} \right) }^
      {1/3}}{6\cdot
     2^{1/3}\,\nu }.$$
     %
The second interesting case is for $s=o$. When  $ w>>0$ the wave
function takes the form
\begin{equation}\label{c}
    \Psi(z,w)= e^{i[\Sigma_{0}z-(\Sigma_{0}/4)\ln w \pm \lambda w]}
\end{equation}
and the corresponding tomogram is
     \beqa\label{st6}
     \mathcal{W} (X_{1},X_{2}, \mu_{1},\mu_{2},\nu_{1},\nu_{2})\propto\frac{1}{|\mu_{1}|}
      &&\left|
      \frac{\exp\left(i\left(-\frac{\Sigma_{0}}{4}\ln\beta_{1}\pm\tilde{\lambda}\beta_{1}\right)\right)}
      {\left(\left(\Sigma_{0}\nu_{2}/4\beta_{1}^{2}\right)+\mu_{2}\right)^{1/2}}+\right.\\ \nonumber
      &&\left.+\frac{\exp \left(i\left(-\frac{\Sigma_{0}}{4}\ln\beta_{2}\pm\tilde{\lambda}\beta_{2}\right)\right)}
      {  \left(\left(\Sigma_{0}\nu_{2}/4\beta_{2}^{2}\right)+\mu_{2}\right)^{1/2} }
      \right|^{2}
 \eeqa
 which can be rewritten in the form
\beqa
 \mathcal{W} (X_{1},X_{2}, \mu_{1},\mu_{2},\nu_{1},\nu_{2})&\propto&\frac{1}{|\mu_{1}|}
 \left(\frac{1}{\left(\left(\Sigma_{0}\nu_{2}/4\beta_{1}^{2}\right)+\mu_{2}\right)}+\right.\\
 \nonumber
 &&+\frac{1}{\left(\left(\Sigma_{0}\nu_{2}/4\beta_{2}^{2}\right)+\mu_{2}\right)}+ \\\nonumber &+&\left. 2 \frac{\cos(\beta_{1}-\beta_{2})}
 {\left(\left(\Sigma_{0}\nu_{2}/4\beta_{1}^{2}\right)+\mu_{2}\right)^{1/2}\left(\left(\Sigma_{0}\nu_{2}/4\beta_{2}^{2}
 \right)+\mu_{2}\right)^{1/2}}\right)\eeqa
where $\beta_{1}$ and $\beta_{2}$ are the solutions of equation
\begin{equation}\label{st7}
    \frac{\partial}{\partial w}
    [-(\Sigma_{0}/4)\ln w \pm \lambda w+\frac{\mu_{2}}{2\nu_{2}}w^{2}-\frac{X_{2}}{\nu_{2}}w]
    =-(\Sigma_{0}/4) \frac{1}{w} \pm \lambda  +\frac{\mu_{2}}{ \nu_{2}}w
    -\frac{X_{2}}{\nu_{2}} =0
\end{equation}

$$\beta_{1}=-\frac{(\pm\tilde{\lambda}-X_{2})-\sqrt{(\pm\tilde{\lambda}\nu_{2}-X_{2})^{2}-\Sigma_{0}\mu_{2}\nu_{2}}}{2\mu_{2}},$$
$$\beta_{2}=-\frac{(\pm\tilde{\lambda}-X_{2})+\sqrt{(\pm\tilde{\lambda}\nu_{2}-X_{2})^{2}-\Sigma_{0}\mu_{2}\nu_{2}}}{2\mu_{2}}.$$
In general, if we are able to achieve the WDW wave function in an
explicit form it is always possible to construct the tomographic
counterpart.

At this point, we have to make a remark. If one takes the wave
function $\Psi(x)$ in the form of the De Broglie wave
${\displaystyle \frac{1}{\sqrt{\pi}}e^{ikx}}$, the probability
density $|\Psi(x)|^2$ is ${\displaystyle \frac{1}{2\pi}}$. Also
the tomogram calculated for this wave function is proportional to
the value ${\displaystyle \frac{1}{|\mu|}}$. Using the tomogram
for reconstructing the wave function, we get divergent integrals.
Due to this fact,one can use a "regularized" wave packet
${\displaystyle \tilde{\Psi}(x)}=A(x)e^{ikx}$ with the amplitude
$A(x)$, e.g., of the Gaussian form ${\displaystyle \sim e^{-g
x^2}}$. In this case, the tomogram assumes the form of a Gaussian
distribution and the density matrix for the pure state
$\tilde{\psi}(x)\tilde{\psi}^*(x)$ can be reconstructed since all
the integrals do diverge. The "regularized" wave packet, having a
highly oscillatory behavior for $g\rightarrow 0$, corresponds to
the Hartle criterion and then to  the presence of Noether
symmetries. In summary, when we have the tomograms as  in the
above examples containing $|\mu|$ in the denominator, the
reconstruction formula can be explained in the described sense.

\section{Discussion and conclusions}
In this paper, we discussed the realization of minisuperspace
Quantum Cosmology adopting a tomographic probability
representation for the wave function of the universe. The physical
meaning of the results is recovered if, in semiclassical WKB
approximation, it is possible to select conserved quantities
(stationary phases) where the wave function is peaked. In this
case, the interpretative Hartle criterion can be applied and
classical trajectories, corresponding to observable universes, are
recovered.  As a matter of fact the existence of Noether
symmetries lead to oscillating components of the wave function,
which present the picks required for a transition from the quantum
cosmological states to the classical universes. In our approach,
we found that a Noether symmetry $\Sigma_0$ implies a factor of
the form $1/|\mu|$ in the tomogram. Noteworthy the tomograms
assume a form which is very close to classical ones, showing in a
natural way the transition from the quantum initial stages of the
universe to its classical evolution. This fact is crucial in our
discussion since tomograms can allow, in principle, a full
description of the universe from its initial quantum states up to
the today observed. Specifically, considering the above considered
models, it is easy to find out,   either from tomograms or from
Noether constants,  exact classical solutions. In the case of
scalar-tensor models, we get
 \begin{eqnarray}
 w(t)&=&[k_1t+k_2]^{3/(s+3)}\,{,}\label{67} \\
 z(t)&=&[k_1t+k_2]^{(s+6)/(s+3)}+z_0\,{,}\label{68}
 \end{eqnarray}
which can be immediately translated into the original
configuration space ${\cal Q}\equiv\{a, \varphi\}$, that is
 \begin{eqnarray}
 a(t)&=&a_0(t-t_0)^{b(s)}\,{,} \label{69} \\
 \varphi(t)&=&\varphi_0(t-t_0)^{q(s)}\,{,} \label{70}
 \end{eqnarray}
 where
 \begin{equation}\label{71}
 b(s)=\frac{2s^2+9s+6}{s(s+3)}\,{,}\quad q(s)=-\frac{2s+3}{s}\,{.}
 \end{equation}
Depending on the value of $s$, we get Friedman, power--law, or
pole-like behaviors, that is all the standard classical
cosmological behaviors. Analogously, for the fourth-order models,
we have, for $s=0$,
\begin{equation}\label{101}
 a(t)=a_0e^{(\lambda/6)t}\,\exp{\left\{-\frac{z_1}{3}\,
 e^{-(2\lambda/3)t}\right\}}\,{,}
 \end{equation}
 \begin{equation}\label{102}
p(t)=p_0e^{(\lambda/6)t}\,\exp{\{z_1\,
 e^{-(2\lambda/3)t}\} }\,{,}
 \end{equation}
where $a_0, p_0$ and $z_1$ are integration constants. It is clear
that $\lambda$ plays the role of a cosmological constant and
inflationary behavior is asymptotically recovered. For $s=-2$, we
get power-law behaviors for $a(t)$ and $p(t)$.

Such solutions, in principle, give rise to "observable universes"
since the set of parameters
$\{H_{0},q_{0},\Omega_{M},\Omega_{\Lambda},\Omega_{k} \}$ can be
obtained, in a standard way, from the solutions $a(t)$,
$\varphi(t)$ and $p(t)$ \cite{peebles}. This fact could be
extremely relevant, also in view of the recent observational
trends which have given rise to the so called {\it Precision
Cosmology} (see for example \cite{wmap,snea,lss}). In fact, the
possibility to trace back the dynamics, via tomograms, from the
today observed parameters up to the initial quantum conditions
could be an interesting approach to formulate comprehensive
cosmological models enclosing early and late evolution.  In other
words, due to this feature, the tomographic approach could be,
with respect to other approaches included the Wigner function one,
the most suitable to allow a connection between inflation and dark
energy cosmology.

To conclude, we point out that all the main results of Quantum
Cosmology can be obtained using, arbitrarily, some well known
formulation of Quantum Mechanics \cite{AmJPhys} including the
density matrix and the Wigner function. This means that, in
principle, also other approaches could be successfully used with
significant results \cite{anastopoulos}. However, as  emphasized
here, the formulation based on  tomographic probability
distribution has some interesting properties which are convenient
to be used as soon as the quantum-classical transitions or the
comparison between classical and quantum pictures are relevant: in
this framework,  tomograms assume a fundamental role being objects
which  fairly describe states both in classical and quantum
domains.

\section*{Acknowledgments}
V.I. Man'ko wants to thank the University of Naples "Federico II"
and the INFN, Sez. di Napoli, for the hospitality.

\end{document}